# Causality and CPT violation from an Abelian Chern–Simons-like term

C. Adam[1], F.R. Klinkhamer[2]

*Institut für Theoretische Physik, Universität Karlsruhe, 76128 Karlsruhe, Germany*

## Abstract

We study a class of generalized Abelian gauge field theories where CPT symmetry is violated by a Chern–Simons-like term which selects a preferred direction in spacetime. Such Chern–Simons-like terms may either emerge as part of the low-energy effective action of a more fundamental theory or be produced by chiral anomalies over a nonsimply connected spacetime manifold. Specifically, we investigate the issues of unitarity and causality. We find that the behaviour of these gauge field theories depends on whether the preferred direction is spacelike or timelike. For a purely spacelike preferred direction, a well-behaved Feynman propagator exists and microcausality holds, which indicates the possibility of a consistent quantization of the theory. For timelike preferred directions, unitarity or causality is violated and a consistent quantization does not seem to be possible.

---

[1]E-mail address: adam@particle.uni-karlsruhe.de

[2]E-mail address: frans.klinkhamer@physik.uni-karlsruhe.de

# 1 Introduction

Lorentz and CPT invariance are two of the cornerstones of modern quantum field theory. Both invariances are respected by the Standard Model of known elementary particles (quarks and leptons) and their interactions. Possible signals of Lorentz and CPT violation could, therefore, be indicative of new physics, e.g. quantum gravity [1, 2] or superstrings [3]. But even within local quantum field theory an anomalous breaking of Lorentz and CPT symmetry might occur, at least for a nontrivial global spacetime structure [4, 5].

Consequently, a considerable amount of attention has been devoted over the last years to the possible occurrence of Lorentz and CPT noninvariance. Phenomenological consequences of breaking Lorentz and CPT symmetry in electromagnetism were studied in Ref. [6]. It was shown that the symmetry breaking would result in, for example, optical activity (birefringence) of the vacuum, that is, a direction-dependent rotation of the linear polarization of an electromagnetic plane wave. Reference [7], in turn, investigated CPT- and Lorentz-noninvariant extensions of the Standard Model (interpreted as low-energy limits of more fundamental theories). Furthermore, there have been extensive discussions in the literature on the possibility of CPT- and Lorentz-symmetry breaking in the gauge field sector induced by radiative corrections of an explicitly symmetry-breaking matter sector, see Refs. [7]–[10] and references therein.

At this point, the question arises whether or not a quantum field theory with Lorentz- and CPT-violating terms can be consistent at all, cf. Refs. [11]–[13]. Also, in each of the papers quoted in the previous paragraph, the CPT- and Lorentz-noninvariant terms in the gauge field sector were of the Chern–Simons type [14]. In this paper, therefore, we intend to study possible implications of a Chern–Simons-like term for the quantization of Abelian gauge fields, focusing on the issues of unitarity and causality.

We start from the following Lagrangian density:

$$\mathcal{L}(x) = -\tfrac{1}{4} F_{\mu\nu}(x) F^{\mu\nu}(x) - \tfrac{1}{2} \xi^{-1} \left(n_\mu A^\mu(x)\right)^2 + \mathcal{L}_{\text{CS–like}}(x) \,, \tag{1.1}$$

with the Chern–Simons-like term

$$\mathcal{L}_{\text{CS–like}}(x) = \tfrac{1}{4} m \, k_\mu \, \epsilon^{\mu\nu\rho\sigma} A_\nu(x) F_{\rho\sigma}(x) \,, \tag{1.2}$$

in terms of the Abelian gauge potential $A_\mu(x)$ and field strength tensor $F_{\mu\nu}(x) \equiv \partial_\mu A_\nu(x) - \partial_\nu A_\mu(x)$. The spacetime metric is taken to have Lorentzian signature $(-, +, +, +)$ and $\epsilon_{\mu\nu\rho\sigma}$ is the completely antisymmetric Levi-Civita symbol, normalized to $\epsilon_{0123} = +1$. (Our conventions, with $\hbar = c = 1$, will be given in more detail later on.)

The Abelian Chern–Simons-like term (1.2) is characterized by a real mass parameter $m$ and a real symmetry-breaking "vector" $k_\mu$ of unit length, which may be spacelike ($k^2 = +1$) or timelike ($k^2 = -1$) but is fixed once and for all (hence, the quotation marks around the word vector). Strictly speaking, $k_\mu$ can also be "lightlike" ($k^2 = 0$), but the present paper considers only the extreme cases, spacelike or timelike $k_\mu$. As long as $k_\mu$ and $m \neq 0$ are fixed external parameters (coupling constants), both Lorentz and CPT invariance are broken, but translation invariance still holds. Note that the Lagrangian term (1.2) is called Chern–Simons-like, because a genuine topological Chern–Simons term exists only in an odd number of dimensions [14].

For later convenience, we have added a gauge-fixing term to the Lagrangian (1.1), where $n_\mu$ determines the axial gauge condition and $\xi$ is a gauge parameter. Choosing an axial gauge, which selects a particular direction $n_\mu$, seems natural because the "vector" $k_\mu$ already selects a preferred direction. In other words, there is no compelling reason to prefer Lorentz-covariant gauge choices over noncovariant ones, cf. Ref. [15].

The Lagrangian (1.1) is Abelian and, therefore, describes a photon-like gauge field. But Eq. (1.1) may as well be interpreted as one component of the quadratic part of a non-Abelian Lagrangian. The discussion that follows is, in principle, also relevant for Lorentz- and CPT-symmetry breaking in a non-Abelian context. Still, the issue of locality may be more subtle for the non-Abelian case due to gauge invariance, as discussed in Section 4 of Ref. [4].

Let us now give in more detail the reasons for studying the Maxwell–Chern–Simons theory (1.1), with broken Lorentz and CPT symmetry. First, a nonzero mass scale $m$ may be introduced by hand as a symmetry-breaking parameter. Possible physical consequences and experimental bounds on the value of $m$ may be studied, as was done in Ref. [6], under the assumption that the Lagrangian (1.1) describes the photon. Recently, it has also been claimed [16] that certain astronomical observations indicate a nonzero value of $m$ for the case of a spacelike Chern–Simons parameter $k_\mu$, but this claim is apparently not substantiated by more accurate data (see Ref. [17] and references therein).

Second, the symmetry-breaking term in the Lagrangian (1.1) may be thought of as being part of the effective action which results from integrating out the fermionic matter fields. Here, the source of the symmetry breaking might be an explicit symmetry-breaking term in the fermionic matter sector [7]–[10]. Alternatively, the symmetry-breaking term in the effective action might be traced to a quantum anomaly which occurs when Weyl fermions in suitable representations are quantized on a nonsimply connected spacetime manifold (e.g. $\mathbb{R}^3 \times S^1$). This CPT anomaly was discovered and described in Ref. [4], where the precise conditions for its occurrence can be found. In this case, the experimentally required smallness of $m$ for photons is naturally accounted for, because the mass scale $m$ is inversely proportional to the linear extension ($L$) of the universe in the compact direction,

$$m_{\text{CPT anomaly}} \sim \alpha \, \hbar \, (L \, c)^{-1} \,, \tag{1.3}$$

with $\alpha$ the fine-structure constant and the dependence on $\hbar$ and $c$ made explicit. For $L \sim 1.5 \, 10^{10}$ lightyears, this mass scale corresponds to $10^{-35}$ eV, which might be within reach of future astronomical observations (this point will be discussed further in Section 6).

Third, the Lagrangian (1.1) may be interpreted as the quadratic gauge field part of a low-energy effective action of a truly fundamental theory, which could, for example, replace point-particles by superstrings, cf. Ref. [3].

Our paper is organized as follows. In Section 2, we focus on the classical aspects of the Maxwell–Chern–Simons theory (1.1) and discuss the resulting dispersion relations and causality behaviour. This turns out to be rather different for spacelike and timelike Chern–Simons parameter $k_\mu$. In Section 3, the Feynman propagator for the Lagrangian (1.1) is calculated both for Minkowskian and Euclidean spacetime. Again, the cases of "spacelike" and "timelike" $k_\mu$ have to be discussed separately. In Section 4, we address the related issue of reflection positivity for the Euclidean theory corresponding to Eq. (1.1), which also depends on the type of parameter $k_\mu$. In Section 5, we determine the field commutators of the quantum field theory based on Eq. (1.1), first for a purely spacelike $k_\mu$. We find that the usual microcausality holds for this case, which is perhaps the most important result of this paper. (Some details of our calculation are relegated to Appendix A.) On the other hand, unitarity and microcausality cannot be maintained simultaneously for a timelike $k_\mu$. In Section 6, finally, we summarize our results and briefly discuss possible applications and open questions. The present paper is, by necessity, quite technical and the general reader may wish to concentrate on Sects. 2 and 6.

## 2   Dispersion relations

As a first step we discuss the dispersion relations which result from the Lagrangian (1.1) without the gauge fixing term and investigate the implications for the causal behaviour of the classical theory. Throughout this section, we take the spacetime manifold $M = \mathbb{R}^4$ and Minkowskian spacetime metric $g_{\mu\nu} = \text{diag}\,(-1, 1, 1, 1)$, with indices running over 0, 1, 2, 3.

The Lagrangian (1.1) then leads to the following dispersion relation for the gauge fields [6]:

$$p^4 + m^2 \left( k^2 p^2 - (k \cdot p)^2 \right) = 0 ,  \qquad (2.1)$$

for momentum $p_\mu = (p_0, p_1, p_2, p_3)$ and $c = 1$.[3] Due to the breaking of Lorentz invariance, there exist preferred coordinate systems. A particular preferred coordinate system for spacelike Chern–Simons parameter $k_\mu$ is one in which $k_\mu$ is purely spacelike ($k_0 = 0$), which we shall choose in the sequel.

Let us discuss this last point in somewhat more detail (see also Ref. [7]). As mentioned in the Introduction, the Chern–Simons parameters $k_\mu$ are considered to be fixed coupling constants (four real numbers) belonging to a particular coordinate system. For localized gauge fields (that is, $A_\mu(x) = 0$ for $|x| \geq R$), one can nevertheless make a Lorentz transformation $x^\mu \to x'^\mu = \Lambda^\mu{}_\lambda x^\lambda + a^\mu$, so that the Chern–Simons-like term (1.2) changes into

$$\tfrac{1}{4}\, m\, k_\mu\, \Lambda^\mu{}_\lambda\, \epsilon^{\lambda\nu\rho\sigma} A_\nu(x') F_{\rho\sigma}(x') \equiv \tfrac{1}{4}\, m\, k'_\lambda\, \epsilon^{\lambda\nu\rho\sigma} A_\nu(x') F_{\rho\sigma}(x') . \qquad (2.2)$$

The new reference frame (with coordinates $x'^\mu$) thus has its own Chern–Simons parameters $k'_\mu$, determined by the old $k_\mu$ and the Lorentz parameters for the change of frame. It is, however, not at all obvious that this change of $k_\mu$ parameters is unitarily implementable for the quantum theory. The quantization of the Maxwell–Chern–Simons theory (1.1) is, therefore, considered rather explicitly in the following sections. For now, we continue our discussion of the classical dispersion relation.

---

[3]For the moment, the velocity parameter $c$ is only used to define the Minkowski spacetime coordinates $(x^0 \equiv ct, x^1, x^2, x^3)$. As will become clear later, $c$ corresponds to the front velocity of light propagation *in vacuo* for the electromagnetic theory based on the Lagrangian (1.1) with a spacelike parameter $k_\mu$.

For a purely spacelike Chern–Simons parameter $k_\mu = (0, \vec{k})$ with $|\vec{k}|^2 = 1$, Eq. (2.1) is a quadratic equation in $p_0^2$, with the following solutions:

$$p_0^2 = |\vec{p}|^2 + \tfrac{1}{2} m^2 \pm \tfrac{1}{2} m \sqrt{m^2 + 4\,(\vec{p}\cdot\vec{k})^2}\,. \tag{2.3}$$

Apparently, there are two very different degrees of freedom, especially towards the infrared ($|\vec{p}| \lesssim m$). (The identification of these two degrees of freedom with circular polarization modes depends on the sign of $\vec{p}\cdot\vec{k}$, cf. Eq. (26) of Ref. [6].) For directions $\vec{p}$ perpendicular to $\vec{k}$, Eq. (2.3) effectively describes one massive degree of freedom with mass $m$, corresponding to the plus sign, and one massless degree of freedom, corresponding to the minus sign. In the $\vec{p}$ direction parallel to $\vec{k}$, both the massive and the massless dispersion relations get distorted. However, both degrees of freedom may still be separated into positive and negative frequency parts, as is obvious from Eq. (2.3). In Fig. 1 we plot the dispersion relation (2.3) restricted to the $(p_0, p_3)$ plane, where $k_\mu$ is assumed to point into the $x^3$ direction as well. The separation into positive and negative frequency parts is clearly seen in Fig. 1.

Without loss of generality we now assume that $k_\mu$ points into the $x^3$ direction, i.e. $k_\mu = (0, 0, 0, 1)$. The dispersion relations for the two degrees of freedom then read

$$p_0^2 = \omega_\pm^2 \equiv p_1^2 + p_2^2 + \widetilde{\omega}_\pm^2\,, \tag{2.4}$$

with

$$\widetilde{\omega}_\pm \equiv \tfrac{1}{2}\left(\sqrt{4 p_3^2 + m^2} \pm m\right)\,. \tag{2.5}$$

Consider, for simplicity, the case of $p_1 = p_2 = 0$ and $p_3 \geq 0$. Then, the dispersion relations $p_0^2 = \widetilde{\omega}_\pm^2$ lead to the phase velocities

$$v_{\text{ph}}^\pm = \frac{\widetilde{\omega}_\pm}{p_3} = \frac{\sqrt{4p_3^2 + m^2} \pm m}{2 p_3} \tag{2.6}$$

and group velocity

$$v_{\text{g}} = \frac{\mathrm{d}\widetilde{\omega}_\pm}{\mathrm{d}p_3} = \frac{2 p_3}{\sqrt{4 p_3^2 + m^2}} \leq 1\,. \tag{2.7}$$

For the case considered and $m \neq 0$, both velocities approach 1 in the limit $p_3 \to \infty$. More generally, the front velocity $v_{\text{f}} \equiv \lim_{|\vec{p}|\to\infty} |\vec{v}_{\text{ph}}|$, which is relevant for signal propagation [18], has the same value 1 in all directions (recall that $c = 1$ in our units).

This classical reasoning already indicates that the causal structure of the theory remains unaffected by the additional CPT-violating term in Eq. (1.1), at least for the case $k_\mu = (0, 0, 0, 1)$. In Section 5.1, we shall find further evidence for this statement by calculating the commutators of the quantized fields.

Before closing this section, we want to contrast the discussion above with that for the case of a timelike Chern–Simons parameter $k_\mu$, which has already been studied in detail by the authors of Ref. [6]. Here, a particular preferred coordinate system is one where $k_\mu$ is purely timelike, $k_\mu = (1, 0, 0, 0)$, which we assume in the following. Again, Eq. (2.1) leads to a quadratic equation in $p_0^2$, with the solutions

$$p_0^2 = \omega_\pm^2 \equiv |\vec{p}|^2 \pm m|\vec{p}| \ . \tag{2.8}$$

(These two degrees of freedom correspond to circular polarization modes, cf. Eq. (26) of Ref. [6].) The dispersion relation (2.8) is plotted in Fig. 2. It is obvious that there is no separation into positive and negative frequency parts.[4] Worse, the energy becomes imaginary at low momenta $|\vec{p}| < m$ for the minus sign in Eq. (2.8). In addition, the group velocities of both degrees of freedom may become arbitrarily large. For the minus sign in Eq. (2.8), one has, for example,

$$\frac{d\omega_-}{d|\vec{p}|} = \frac{2|\vec{p}| - m}{2\sqrt{|\vec{p}|^2 - m|\vec{p}|}} \ , \tag{2.9}$$

which is singular at $|\vec{p}| = m$. These results indicate that the case of timelike Chern–Simons parameter $k_\mu$ is rather different from the case of spacelike $k_\mu$ and does not allow for quantization. In the next section, we shall find further evidence for this statement by investigating the Feynman propagator.

## 3 Feynman propagator in a general axial gauge

We now consider the Feynman propagator which may be formally derived from the Lagrangian (1.1), and investigate what may be inferred for the possible quantization of the theory.

The action corresponding to Eq. (1.1) can be re-expressed as follows:

$$S = \tfrac{1}{2} \int_{\mathbb{R}^4} d^4x \, A_\mu \left( g^{\mu\nu} \Box - \partial^\mu \partial^\nu - (1/\xi) \, n^\mu n^\nu - m \, \epsilon^{\mu\nu\rho\sigma} k_\rho \partial_\sigma \right) A_\nu \ , \tag{3.1}$$

---

[4]This fundamental difference of the Maxwell–Chern–Simons theory (1.1) for the case of, say, $k_\mu = (1, 0, 0, 0)$ and $k_\mu = (0, 0, 0, 1)$ traces back to our fixed choice of the time coordinate, namely $x^0 \equiv ct$.

so that the inverse propagator in momentum space becomes

$$\left(G^{-1}\right)^{\mu\nu}(p) = -p^2 g^{\mu\nu} + p^\mu p^\nu - (1/\xi) n^\mu n^\nu - i\, m\, \epsilon^{\mu\nu\rho\sigma} k_\rho p_\sigma \,. \tag{3.2}$$

The corresponding propagator, which obeys $(G^{-1})^{\mu\nu} G_{\nu\lambda} = -\delta^\mu_\lambda$, reads

$$\begin{aligned}
G_{\nu\lambda}(p) &= \left(g_{\nu\lambda} + \frac{\xi\, p^2 + n^2 + (m^2/p^2)(k\cdot n)^2 + \xi\, m^2 k^2 - \xi\,(m^2/p^2)(k\cdot p)^2}{(p\cdot n)^2} p_\nu p_\lambda \right. \\
&\quad - \frac{1}{(p\cdot n)} (p_\nu n_\lambda + n_\nu p_\lambda) + \frac{m^2}{p^2} k_\nu k_\lambda - \frac{m^2 (k\cdot n)}{p^2(p\cdot n)} (p_\nu k_\lambda + k_\nu p_\lambda) \\
&\quad \left. + i\, m\, \epsilon_{\nu\lambda\alpha\beta} \left( \frac{(p\cdot k)}{p^2(p\cdot n)} p^\alpha n^\beta - \frac{1}{(p\cdot n)} k^\alpha n^\beta \right) \right) K \,,
\end{aligned} \tag{3.3}$$

with

$$K \equiv \frac{p^2}{p^4 + m^2 \left(k^2 p^2 - (k\cdot p)^2\right)} \,. \tag{3.4}$$

(Note that the equivalent propagator in a covariant gauge has already been computed in Ref. [19].)

Up till now, the calculation of the propagator was formal and purely algebraic. We did not discuss the pole structure, nor even define whether we are in Minkowski or Euclidean spacetime.[5] A systematic treatment can be given for the spurious singularities related to the axial gauge vector $n_\mu$ (see, for example, Ref. [15]), and we ignore these singularities in the following. Instead, we focus on the pole structure of the propagator function $K$ as given by Eq. (3.4). For clarity, we relabel our previous Chern–Simons parameter $k_\mu$ in Minkowski spacetime as $k_\mu^{\rm M}$ and use $k_\mu^{\rm E}$ in Euclidean space.

First, let us discuss the case of a purely spacelike $k_\mu^{\rm M} = (0,0,0,1)$, with index $\mu$ running over 0, 1, 2, 3. For Minkowski spacetime with metric signature $(-,+,+,+)$, we get

$$K = \frac{-p_0^2 + |\vec{p}|^2}{\left(p_0^2 - |\vec{p}|^2 - \frac{m^2}{2} - \frac{m}{2}\sqrt{m^2 + 4p_3^2} + i\epsilon\right)\left(p_0^2 - |\vec{p}|^2 - \frac{m^2}{2} + \frac{m}{2}\sqrt{m^2 + 4p_3^2} + i\epsilon\right)} \,, \tag{3.5}$$

---

[5]The classical Euclidean theory is derived from Eq. (1.1) by rotating both $x^0$ and $k^0$ to the imaginary axis. (See Ref. [20] for a general discussion of Euclidean field theories.) An alternative Euclidean theory could perhaps be defined by keeping $k^0$ absolutely fixed, but this is not what has been discussed in, for example, Refs. [4, 7].

where both poles are displaced with the help of the usual Feynman $i\epsilon$ prescription ($\epsilon = 0^+$), cf. Refs. [21, 22]. For Euclidean space with metric signature $(+, +, +, +)$ and indices running over 4, 1, 2, 3, we find instead

$$K = \frac{p_4^2 + |\vec{p}|^2}{\left(p_4^2 + |\vec{p}|^2 + \frac{m^2}{2} + \frac{m}{2}\sqrt{m^2 + 4p_3^2}\right)\left(p_4^2 + |\vec{p}|^2 + \frac{m^2}{2} - \frac{m}{2}\sqrt{m^2 + 4p_3^2}\right)} \,. \quad (3.6)$$

The poles from both factors in the denominator are placed on the positive and negative imaginary axis of the complex $p_4$ plane. A Wick rotation [21, 22] to Minkowski spacetime can be performed and leads to the $i\epsilon$ prescription (3.5) for the Feynman propagator (3.3) of both degrees of freedom. Hence, the propagator is well-behaved, at least for the case of a purely spacelike Chern–Simons parameter.

For a purely timelike $k_\mu^\mathrm{M} = (1, 0, 0, 0)$ in Minkowski spacetime, with index $\mu$ running over 0, 1, 2, 3, we obtain

$$\begin{aligned} K &= -\frac{p_0^2 - |\vec{p}|^2}{(p_0^2 - |\vec{p}|^2 + m|\vec{p}|)(p_0^2 - |\vec{p}|^2 - m|\vec{p}|)} \\ &= -\frac{1}{2(p_0^2 - |\vec{p}|^2 + m|\vec{p}|)} - \frac{1}{2(p_0^2 - |\vec{p}|^2 - m|\vec{p}|)} \,. \end{aligned} \quad (3.7)$$

For low momenta $|\vec{p}| < m$, the poles in the first term are placed on the imaginary $p_0$ axis, which means that the energy becomes imaginary. This, in turn, implies that unitarity is violated already at tree level, i.e. for the free theory (1.1). The region $|\vec{p}| < m$ has, therefore, to be excluded for this degree of freedom. (The situation is similar to the case of a tachyon field with dispersion relation $p_0^2 - |\vec{p}|^2 + m^2 = 0$, where the region $|\vec{p}| < m$ has to be excluded in order to maintain unitarity of the quantum field theory at tree level. See, for example, the discussion in Refs. [23, 24].) But we shall find in Section 5.2 that exclusion of the region $|\vec{p}| < m$ leads to a violation of microcausality.

If we now assume a purely "timelike" $k_\mu^\mathrm{E} = (1, 0, 0, 0)$ in Euclidean space, with index $\mu$ running over 4, 1, 2, 3, the function $K$ becomes

$$\begin{aligned} K &= \frac{p_4^2 + |\vec{p}|^2}{(p_4^2 + |\vec{p}|^2 + im|\vec{p}|)(p_4^2 + |\vec{p}|^2 - im|\vec{p}|)} \\ &= \frac{1}{2(p_4^2 + |\vec{p}|^2 + im|\vec{p}|)} + \frac{1}{2(p_4^2 + |\vec{p}|^2 - im|\vec{p}|)} \,. \end{aligned} \quad (3.8)$$

Here, the poles of the first (second) term are placed in the second and fourth (first and third) quadrants of the complex $p_4$ plane. In order to determine the behaviour of the propagator (3.8) under a Wick rotation to Minkowski space, we have to remember that according to our prescription we have to rotate $k_4$ as well. This makes that the poles of Eq. (3.8) move under Wick rotation. For sufficiently small $|\vec{p}|$, two poles will, in fact, move to the real axis and, therefore, cross the Wick-rotated $p_4$-axis. In short, the analytic behaviour of the propagator is problematic for the case of a purely timelike Chern–Simons parameter.

For spacelike $k_\mu$ in Minkowski spacetime, we have used up till now a special coordinate system in which $k_\mu$ is purely spacelike, that is, $k_0 = 0$ exactly. Let us, finally, relax this condition and investigate what happens if we allow for $k_0 \neq 0$. In general, the four roots of the denominator of Eq. (3.4) are rather complicated. We shall, therefore, make some simplifying assumptions. By choosing $k_1 = k_2 = 0$, we can restrict ourselves to the plane $p_1 = p_2 = 0$. Also, we choose units of energy and momentum such that $\tilde{k}_\mu \equiv mk_\mu = (\tilde{k}_0, 0, 0, 1)$. With these assumptions, we still find four real roots $p_0 = r_i$, $i = 1 \ldots 4$, as long as $|\tilde{k}_0| < 1$.

With the same simplifications, we find in Euclidean space the following four roots $p_4 = q_i$, $i = 1 \ldots 4$, for the denominator of Eq. (3.4):

$$q_1 = \frac{i}{2}\left(1 - \sqrt{1 + 4i\tilde{k}_4 p_3 + 4p_3^2}\right), \quad q_2 = q_1^*,$$

$$q_3 = \frac{i}{2}\left(1 + \sqrt{1 + 4i\tilde{k}_4 p_3 + 4p_3^2}\right), \quad q_4 = q_3^*. \tag{3.9}$$

For $\tilde{k}_4 \neq 0$, $q_1$ and $q_3$ have nonzero real parts of opposite sign ($\operatorname{Re} q_1 = -\operatorname{Re} q_3$) and the four poles of Eq. (3.4) are placed in all four quadrants of the complex $p_4$ plane. Under a Wick rotation, with $k_4$ rotated as well, all four poles (3.9) move towards the imaginary axis together with $p_4$ (as long as $\tilde{k}_\mu$ is "spacelike," $|\tilde{k}_4| < 1$), and a Wick rotation may be performed without crossing poles in the complex $p_4$ plane. Hence, the propagator is well-behaved, provided the Chern–Simons parameter is spacelike.

This completes our elementary discussion of the Feynman propagator for the Maxwell–Chern–Simons theory (1.1). In the next section, we will study the Euclidean propagator in somewhat more detail.

## 4   Reflection positivity

An important condition for the quantization of a field theory in the Euclidean formulation is reflection positivity [25, 20]. This condition is essential for establishing the existence of a positive semi-definite self-adjoint Hamiltonian $H$ in Minkowski spacetime, with the corresponding unitary time evolution operator $\exp(-iHt)$.

The reflection positivity condition for an Euclidean two-point function is simply

$$\langle \Theta \left( \phi(x^4, \vec{x}) \right) \phi(x^4, \vec{x}) \rangle \geq 0 \,, \tag{4.1}$$

where $x^4$ is the Euclidean time coordinate, $\phi(x^4, \vec{x})$ a scalar field of the theory, and $\Theta : \phi(x^4, \vec{x}) \to \phi^\dagger(-x^4, \vec{x})$ the reflection operation. Reflection positivity then gives the following inequality for the scalar Euclidean propagator function $G(p_4, \vec{p})$:

$$\int d^3p \int_{-\infty}^{\infty} dp_4 \, e^{-ip_4 x^4} G(p_4, \vec{p}) \equiv \int d^3p \, G(x^4, \vec{p}) \geq 0 \,, \tag{4.2}$$

for arbitrary values of $x^4$. By choosing suitable smearing functions, it is even possible to derive the stronger condition $G(x^4, \vec{p}) \geq 0$, but for our purpose the condition (4.2) suffices.

For the gauge-invariant degrees of freedom of the Maxwell–Chern–Simons theory (1.1), it turns out to be sufficient to check the issue of reflection positivity for the Euclidean propagator function $K$ (as introduced in Eq. (3.3) above), thereby effectively reducing the problem to the investigation of a scalar two-point function. Concretely, we then have to verify whether or not the inequality (4.2) holds for our propagator function $K(p_4, \vec{p})$.

For the case of purely "spacelike" $k_\mu^E = (0, 0, 0, 1)$, with index $\mu$ running over 4, 1, 2, 3, the function $K(p_4, \vec{p})$ is given by Eq. (3.6) and we get

$$\begin{aligned} K(x^4, \vec{p}) &= \int_{-\infty}^{+\infty} dp_4 \, e^{-ip_4 x^4} \frac{p_4^2 + |\vec{p}|^2}{(p_4^2 + \omega_+^2)(p_4^2 + \omega_-^2)} \\ &= \pi \frac{\omega_+ e^{-\omega_+ |x^4|} - \omega_- e^{-\omega_- |x^4|}}{\omega_+^2 - \omega_-^2} + \pi \frac{|\vec{p}|^2}{\omega_+ \omega_-} \frac{\omega_+ e^{-\omega_- |x^4|} - \omega_- e^{-\omega_+ |x^4|}}{\omega_+^2 - \omega_-^2} \\ &= \pi \frac{(|\vec{p}|^2 - \omega_-^2) \omega_+ e^{-\omega_- |x^4|} + (\omega_+^2 - |\vec{p}|^2) \omega_- e^{-\omega_+ |x^4|}}{\omega_+ \omega_- (\omega_+^2 - \omega_-^2)} \,, \end{aligned} \tag{4.3}$$

where Eqs. (3.728.1) and (3.728.3) of Ref. [26] have been used to evaluate the integral and the frequencies $\omega_\pm$ are defined in Eq. (2.4). This expression is manifestly positive semi-definite, since $\omega_- \leq |\vec{p}| \leq \omega_+$, and reflection positivity (4.2) holds.

For the case of purely "timelike" $k_\mu^E = (1, 0, 0, 0)$, the function $K(p_4, \vec{p})$ is given by Eq. (3.8) and we get

$$\begin{aligned} K(x^4, \vec{p}) &= \int_{-\infty}^{+\infty} dp_4 \, e^{-ip_4 x^4} \frac{p_4^2 + |\vec{p}|^2}{p_4^4 + 2p_4^2 |\vec{p}|^2 + |\vec{p}|^4 + m^2 |\vec{p}|^2} \\ &= \int_0^{+\infty} dp_4 \, 2\cos(p_4 x^4) \frac{p_4^2 + |\vec{p}|^2}{p_4^4 + 2p_4^2 b^2 \cos 2a + b^4} \,, \end{aligned} \tag{4.4}$$

with

$$\cos 2a \equiv |\vec{p}|^2/b^2\,, \quad b^2 \equiv |\vec{p}|\sqrt{|\vec{p}|^2 + m^2}\,.$$

The integration over $p_4$ can be performed explicitly and we obtain, using Eqs. (3.733.1) and (3.733.3) of Ref. [26],

$$\begin{aligned}
K(x^4, \vec{p}) &= \pi \exp(-|x^4|\,b\cos a) \left( \frac{\sin[a - |x^4|\,b\sin a]}{b \sin 2a} + \frac{|\vec{p}|^2}{b^2} \frac{\sin[a + |x^4|\,b\sin a]}{b \sin 2a} \right) \\
&= \pi \exp(-|x^4|\,b\cos a)\,\cos[a + |x^4|\,b\sin a]/b\,.
\end{aligned} \quad (4.5)$$

Clearly, this expression is not positive semi-definite (as long as $m \neq 0$) and numerical integration over $\vec{p}$ shows reflection positivity (4.2) to be violated for large enough values of $m|x^4|$. The different behaviour of, respectively, Eqs. (4.3) and (4.5) is caused by the different pole structure in Eqs. (3.6) and (3.8), which was also the crux of the previous section.

## 5 Microcausality

Having dealt with unitarity, we continue our investigation of the hypothetical quantum field theory based on the Lagrangian (1.1) and focus on the issue of causality. Minkowskian conventions, with metric signature $(-, +, +, +)$ and indices running over 0, 1, 2, 3, are assumed throughout this section and the units are such that $c = \hbar = 1$.

### 5.1 Purely spacelike Chern–Simons parameter

Let us, again, start with the case of a purely spacelike "vector" $k_\mu = (0, \vec{k})$. We prefer to use a physical gauge condition, in order to avoid the problem of constructing the subspace of physical states. Furthermore, we will try to connect to the well-known results of Quantum Electrodynamics, i.e. the Lagrangian (1.1) for $m = 0$. We, therefore, switch from the general axial gauge to the Coulomb gauge, $\vec{\partial} \cdot \vec{A} = 0$, cf. Refs. [22, 27].

The resulting commutator for the gauge field $\vec{A}(x^0, \vec{x})$ is then given by

$$[A_i(x), A_j(0)] = i\,T_{ij}(-i\partial_0, -i\vec{\partial})\,D(x)\,, \quad (5.1)$$

with the commutator function

$$D(x) = (2\pi)^{-4} \oint_C dp_0 \int d^3p\, \frac{e^{ip_0 x^0 + i\vec{p}\cdot\vec{x}}}{(p^2)^2 + m^2\left(p^2|\vec{k}|^2 - (\vec{p}\cdot\vec{k})^2\right)}\,, \quad (5.2)$$

for an integration contour $C$ that encircles all four poles of the integrand in the counterclockwise direction, cf. Appendix A1 of Ref. [28]. The denominator of Eq. (5.2) is given by the dispersion relation (2.1) for purely spacelike Chern–Simons parameter $k_\mu$. The "tensor" $T_{ij}$ on the right-hand side of the commutation relation (5.1) is found to be given by

$$T_{ij}(p_0, \vec{p}) = (p_0^2 - |\vec{p}|^2)\, \pi_{ij} - m^2\, s_{ij} + imp_0\, a_{ij} , \qquad (5.3)$$

with

$$\pi_{ij} \equiv \delta_{ij} - \frac{p_i p_j}{|\vec{p}|^2} , \qquad (5.4)$$

$$s_{ij} \equiv \left(k_i - \frac{\vec{p}\cdot\vec{k}}{|\vec{p}|^2} p_i\right)\left(k_j - \frac{\vec{p}\cdot\vec{k}}{|\vec{p}|^2} p_j\right) , \qquad (5.5)$$

$$a_{ij} \equiv \epsilon_{ija}k_a + \frac{p_i}{|\vec{p}|^2} \epsilon_{jab}p_a k_b - \frac{p_j}{|\vec{p}|^2} \epsilon_{iab}p_a k_b = \frac{\vec{p}\cdot\vec{k}}{|\vec{p}|^2} \epsilon_{ijl}p_l . \qquad (5.6)$$

One immediately verifies that the commutator (5.1) respects the Coulomb gauge, since $p_i\, T_{ij} = 0$ and $p_j\, T_{ij} = 0$. Further details on the derivation of this commutation relation can be found in Appendix A.

Microcausality (i.e. the commutativity of local observables with spacelike separations, cf. Refs. [12, 21, 28]) holds, provided that:

1. the commutator function $D(x)$ vanishes for spacelike separations $|x^0| < |\vec{x}|$, and

2. the poles of the type $|\vec{p}|^{-2}$ which occur in the "tensor" $T_{ij}$ are absent in the commutators of physical, gauge-invariant fields (i.e. the electric and magnetic fields).

In our case, the commutators of the electric field $\vec{E} \equiv \partial_0 \vec{A} - \vec{\partial} A_0$ and magnetic field $\vec{B} \equiv \vec{\partial} \times \vec{A}$ are found to be the following (see Appendix A for details):

$$[E_i(x), E_j(0)] = \Big((\partial_0^2 - \vec{\partial}^2)(\delta_{ij}\partial_0^2 - \partial_i\partial_j) + m^2\partial_0^2 k_i k_j$$
$$-m\partial_0^3 \epsilon_{ijl}k_l - m\partial_0(\partial_i\epsilon_{jab}\partial_a k_b - \partial_j\epsilon_{iab}\partial_a k_b)\Big) i\, D(x) , \qquad (5.7)$$

$$[E_i(x), B_j(0)] = \Big((\partial_0^2 - \vec{\partial}^2)\epsilon_{ijl}\partial_l\partial_0 + m^2\partial_0 k_i \epsilon_{jab}\partial_a k_b$$
$$+m\partial_0^2(\vec{k}\cdot\vec{\partial})\delta_{ij} - m(\vec{k}\cdot\vec{\partial})\partial_i\partial_j - m(\partial_0^2 - \vec{\partial}^2)\partial_i k_j\Big) i\, D(x) , \qquad (5.8)$$

$$[B_i(x), B_j(0)] = \Big((\partial_0^2 - \vec{\partial}^2)(\delta_{ij}\vec{\partial}^2 - \partial_i\partial_j) + m^2\{\delta_{ij}(\vec{\partial}^2|\vec{k}|^2 - (\vec{k}\cdot\vec{\partial})^2) - |\vec{k}|^2\partial_i\partial_j$$
$$- k_i k_j \vec{\partial}^2 + (\partial_i k_j + k_i \partial_j)(\vec{k}\cdot\vec{\partial})\} - m\partial_0(\vec{k}\cdot\vec{\partial})\epsilon_{ijl}\partial_l\Big)\, i\, D(x)\ . \quad (5.9)$$

Remark that poles of the type $|\vec{p}|^{-2}$, which could spoil causality, are indeed absent in these commutators of physical field operators. In addition, we recover the Jordan–Pauli commutators of Quantum Electrodynamics [27] in the limit $m \to 0$ (remember that our $D(x)$ for $m \to 0$ obeys the massless Klein–Gordon equation squared, $\Box\Box D = 0$).

We still have to discuss the commutator function (5.2). Henceforth, we assume that $\vec{k}$ points into the 3 direction, $\vec{k} = (0, 0, 1)$. We first observe that (5.2) vanishes for equal times ($x^0 = 0$), because of the symmetry in $p_0$ of the integrand (5.2), which results in a cancellation of the residues (compare Eq. (5.10) for $\widetilde{x}^0 = 0$ below). The commutator function is also zero for $(x^0)^2 < (x^1)^2 + (x^2)^2$, because the integrand can be made to be symmetric in a new variable $p_0'$, which is related to $p_0$ by a conventional Lorentz boost involving only $p_0$, $p_1$, and $p_2$. We will now show that the commutator function $D(x)$ vanishes, in fact, over the *whole* spacelike region $|x^0| < |\vec{x}|$. The reader who is not interested in the details may skip the rest of this subsection.

For $(x^0)^2 \geq (x^1)^2 + (x^2)^2$, it is still useful to perform a Lorentz transformation involving only $p_0$, $p_1$ and $p_2$, because there exists a transformation which allows us to rewrite Eq. (5.2) as

$$D(\widetilde{x}^0, x^3) = (2\pi)^{-4} \oint_C dp_0 \int d^3p\, \frac{e^{ip_0\widetilde{x}^0 + ip_3 x^3}}{(p^2)^2 + m^2\left(p^2|\vec{k}|^2 - (\vec{p}\cdot\vec{k})^2\right)},$$

$$= (2\pi)^{-4} \oint_C dp_0 \int d^3p\, \frac{e^{ip_0\widetilde{x}^0 + ip_3 x^3}}{(p_0 - \omega_+)(p_0 + \omega_+)(p_0 - \omega_-)(p_0 + \omega_-)}, \quad (5.10)$$

with

$$\widetilde{x}^0 \equiv \sqrt{\frac{(x^0)^2 - (x^1)^2 - (x^2)^2}{(x^0)^2}}\, x^0\ . \quad (5.11)$$

(For $(x^0)^2 < (x^1)^2 + (x^2)^2$, we can effectively set $\widetilde{x}^0 = 0$.) The contour integral is readily performed,

$$D(\widetilde{x}^0, x^3) = (2\pi)^{-3} \int d^3p\, e^{ip_3 x^3} \left(\frac{\sin\omega_-\widetilde{x}^0}{\omega_-(\omega_+^2 - \omega_-^2)} - \frac{\sin\omega_+\widetilde{x}^0}{\omega_+(\omega_+^2 - \omega_-^2)}\right)\ , \quad (5.12)$$

with the roots $\omega_\pm$ explicitly given by Eq. (2.4). The integral (5.12) obviously vanishes for $\widetilde{x}^0 = 0$ (as long as $x^3 \neq 0$), and we are interested in determining its behavior for other values of $\widetilde{x}^0$. We will start by demonstrating that $D(\widetilde{x}^0, x^3)$ at the time slice $\widetilde{x}^0 = 0$ is ultra-local in $x^3$, i.e. $\partial_{\widetilde{x}^0}^n D(\widetilde{x}^0, x^3)|_{\widetilde{x}^0=0}$ is a sum of derivatives of the delta function $\delta(x^3)$.

If $n$ is even, then $\partial_{\widetilde{x}^0}^n D(\widetilde{x}^0, x^3)|_{\widetilde{x}^0=0}$ is obviously zero. If $n = 2l + 1$ is odd, then one has

$$\partial_{\widetilde{x}^0}^{2l+1} D(\widetilde{x}^0, x^3)|_{\widetilde{x}^0=0} = (-1)^{l+1} (2\pi)^{-3} \int d^3p \, e^{ip_3 x^3} \frac{(\omega_+^2)^l - (\omega_-^2)^l}{\omega_+^2 - \omega_-^2} \,. \tag{5.13}$$

Two remarks are in order. First, the fraction in the integrand is, in fact, a polynomial in $\omega_+^2$ and $\omega_-^2$, namely

$$\frac{(\omega_+^2)^l - (\omega_-^2)^l}{\omega_+^2 - \omega_-^2} = (\omega_+^2)^{l-1} + (\omega_+^2)^{l-2}\omega_- + \ldots + \omega_+^2(\omega_-^2)^{l-2} + (\omega_-^2)^{l-1} \equiv P_{2l-2} \,. \tag{5.14}$$

Second, if we temporarily re-express $\omega_\pm^2$ as

$$\omega_\pm^2 = a \pm b \,, \tag{5.15}$$

where $a$ is a polynomial in the momenta $\vec{p}$ and $b$ is the square-root of a polynomial, then the above polynomial (5.14) only depends on even powers of $b$, $P_{2l-2} = P_{2l-2}(a, b^2)$. The last observation follows from the simple fact that $P_{2l-2}$ is invariant under the interchange $\omega_+^2 \leftrightarrow \omega_-^2$. These two remarks make clear that $P_{2l-2}$ is a polynomial in the momenta $p_1$, $p_2$, and, especially, $p_3$, which implies ultra-locality.

The finite domain of vanishing $D$ may be determined by a direct evaluation of the integral (5.12). The end result of a straightforward calculation is that

$$D(\widetilde{x}^0, x^3) = 0 \,, \quad \text{for} \ |\widetilde{x}^0| < |x^3| \,, \tag{5.16}$$

which corresponds to the usual spacelike region $(x^0)^2 < (x^1)^2 + (x^2)^2 + (x^3)^2$, see Eq. (5.11) above. The calculation proceeds in three steps.

First, one notes that the factors $(\omega_+^2 - \omega_-^2)$ in the denominators of (5.12) are independent of $p_1$ and $p_2$, so that these integrals can be readily performed for $\widetilde{x}^0 \neq 0$,

$$D(\widetilde{x}^0, x^3) = -\frac{1}{4\pi^2 \widetilde{x}^0} \int_{-\infty}^{\infty} dp_3 \, e^{ip_3 x^3} \frac{\cos \widetilde{\omega}_+ \widetilde{x}^0 - \cos \widetilde{\omega}_- \widetilde{x}^0}{\widetilde{\omega}_+^2 - \widetilde{\omega}_-^2} \,, \tag{5.17}$$

where the $\widetilde{\omega}_\pm \equiv \omega_\pm|_{p_1^2+p_2^2=0}$ are given by Eq. (2.5). (To arrive at Eq. (5.17) we have dropped the contribution at $p_1^2 + p_2^2 = \infty$, which corresponds to a rapidly oscillating function of $\widetilde{x}^0$ that vanishes upon integration.) Note that the Taylor expansion of the integrand of (5.17) in powers of $\widetilde{x}^0$ has precisely the polynomials (5.14) as coefficients, but now in terms of $\widetilde{\omega}_\pm$.

Second, we replace the variable $p_3$ by $\phi$, which is defined as follows

$$p_3 \equiv \tfrac{1}{2} m \sinh \phi, \quad \sqrt{p_3^2 + \tfrac{1}{4} m^2} \equiv \tfrac{1}{2} m \cosh \phi. \tag{5.18}$$

This change of variables eliminates the denominator $(\tilde{\omega}_+^2 - \tilde{\omega}_-^2)$ in (5.17), so that only exponentials remain in the integrand. (The same procedure is followed in Section 15.1 of Ref. [29] for the standard commutator function of massive scalars.) The result is then

$$D(\tilde{x}^0, x^3) = -\frac{1}{8\pi^2 m \, \tilde{x}^0} \int_{-\infty}^{\infty} d\phi \, e^{i p_3 x^3} \left( \cos \tilde{\omega}_+ \tilde{x}^0 - \cos \tilde{\omega}_- \tilde{x}^0 \right), \tag{5.19}$$

with $p_3$ and $\tilde{\omega}_\pm$ defined in terms of $\phi$.

Third, the integral over $\phi$ can be evaluated, taking care of the relative signs and magnitudes of $\tilde{x}^0$ and $x_3$. For the case of $0 < \tilde{x}^0 < x^3$, we write

$$\tilde{x}^0 \equiv \sqrt{x^2} \sinh \phi_0, \quad x^3 \equiv \sqrt{x^2} \cosh \phi_0. \tag{5.20}$$

Defining $\mu \equiv \tfrac{1}{2} m \sqrt{x^2}$, the $\phi$ integral in (5.19) becomes after a simple manipulation

$$\int_{-\infty}^{\infty} d\phi \, (i \sin[\mu \sinh \phi_0]) \left( e^{i\mu \sinh(\phi + \phi_0)} - e^{i\mu \sinh(\phi - \phi_0)} \right). \tag{5.21}$$

The first factor in brackets is a constant which can be taken out of the integral. But the remaining integral of the second factor in brackets vanishes trivially (in the second term shift $\phi \to \phi + 2\phi_0$). Since the $\phi$ integral (5.19) is even in both $\tilde{x}^0$ and $x^3$, and the original commutator function $D$ as given by Eq. (5.12) manifestly vanishes for $\tilde{x}^0 = 0$, this establishes the result (5.16) announced above.

It may also be instructive to see what happens for the case of, say, $0 < x^3 < \tilde{x}^0$. Defining

$$\tilde{x}^0 \equiv \sqrt{-x^2} \cosh \phi_0, \quad x^3 \equiv \sqrt{-x^2} \sinh \phi_0, \tag{5.22}$$

one now gets for the $\phi$ integral in Eq. (5.19)

$$\int_{-\infty}^{\infty} d\phi \, (i \sin[\mu \cosh \phi_0]) \left( e^{i\mu \cosh(\phi + \phi_0)} - e^{-i\mu \cosh(\phi - \phi_0)} \right), \tag{5.23}$$

with $\mu \equiv \tfrac{1}{2} m \sqrt{-x^2}$. Taking out the constant factor and making a change of variables $\phi \to \phi + 2\phi_0$ for the second term, one obtains

$$\int_{-\infty}^{\infty} d\phi \left( e^{i\mu \cosh(\phi + \phi_0)} - e^{-i\mu \cosh(\phi + \phi_0)} \right), \tag{5.24}$$

which need not vanish. Using Eq. (8.421) of Ref. [26], the integral (5.24) gives $2\pi i J_0(\mu)$, where the Bessel function $J_0(\mu)$ is, in general, nonzero. All together, one has

$$D(\widetilde{x}^0, x^3) = \frac{1}{8\pi} \epsilon(\widetilde{x}^0) \frac{\sin\left(\frac{1}{2}m\widetilde{x}^0\right)}{\frac{1}{2}m\widetilde{x}^0} J_0\left(\frac{1}{2}m\sqrt{-x^2}\right), \quad \text{for } |\widetilde{x}^0| > |x^3|, \qquad (5.25)$$

where the antisymmetry in $\widetilde{x}^0$ has been made explicit.[6] For fixed timelike separation $x^\mu$ and Chern–Simons mass scale $m \to 0$, the commutator function approaches a constant value $\pm(8\pi)^{-1}$. (Remark that derivatives operating on $D$ will result in further singularities on the null-cone $x^2 = 0$.) This completes our discussion of the gauge field commutator (5.1) for the case of a purely spacelike Chern–Simons parameter $k_\mu = (0, 0, 0, 1)$, with microcausality established.

## 5.2 Purely timelike Chern–Simons parameter

Let us, briefly, discuss the commutator function for the case of a purely timelike "vector" $k_\mu$. In this case there is no invariant separation of the dispersion relation into positive and negative frequency parts (see Fig. 2). However, as Lorentz invariance is broken anyway, we may simply choose to quantize in the particular coordinate frame where $k_\mu = (1, 0, 0, 0)$. Specifically, we want to study the degree of freedom with dispersion relation $p_0^2 = |\vec{p}|^2 - m|\vec{p}|$. For this degree of freedom, the region $|\vec{p}| < m$ has to be excluded in order to maintain unitarity, as was mentioned a few lines below Eq. (3.7). The relevant commutator [29] is then

$$[\Phi(0), \Phi(x)] = i\tilde{D}(x), \qquad (5.26)$$

with $\Phi(x)$ the quantum field corresponding to this particular degree of freedom of the gauge field (recall that $\hbar = c = 1$) and the commutator function

$$\begin{aligned}
\tilde{D}(x) &= \frac{1}{i(2\pi)^3} \int d^4p\, \delta\left(p_0^2 - |\vec{p}|^2 + m|\vec{p}|\right) \theta(|\vec{p}| - m)\, \epsilon(p_0)\, e^{ip\cdot x} \\
&= \frac{1}{(2\pi)^3} \int d^3p\, \frac{\theta(|\vec{p}| - m)}{\sqrt{|\vec{p}|^2 - m|\vec{p}|}} \sin\left(\sqrt{|\vec{p}|^2 - m|\vec{p}|}\, x^0\right) e^{i\vec{x}\cdot\vec{p}}.
\end{aligned} \qquad (5.27)$$

Here, $\epsilon(x) \equiv x/|x|$ and $\theta$ is the usual step function, $\theta(x) = 0$ for $x < 0$ and $\theta(x) = 1$ for $x > 0$. We will now demonstrate that microcausality is violated for this commutator function, i.e. $\tilde{D}(x) \neq 0$ somewhere in the spacelike region $|x^0| < |\vec{x}|$.

---

[6]The structure of our commutator function (5.25) closely resembles the one obtained for a CPT-violating massive Dirac fermion, as given in Appendix E of the first paper in Ref. [7]. Note, however, that in our case $m$ sets the scale of the CPT violation for the photons.

If $\tilde{D}(x)$ were to vanish for $|x^0| < |\vec{x}|$, this would imply that $\mathcal{D}(x^0, x^3) \equiv \int \mathrm{d}x^1 \mathrm{d}x^2 \tilde{D}(x)$ = 0 for $|x^0| < |x^3|$. So, let us show that the latter relation is violated. If $\mathcal{D}(x^0, x^3)$, in turn, were to vanish for $|x^0| < |x^3|$, this would imply that $\partial_0 \mathcal{D}(x^0, x^3)|_{x^0=0}$ had to be an ultra-local expression in $x^3$, but it can be easily checked that this is not the case. Indeed, we calculate

$$\partial_0 \mathcal{D}(x^0, x^3)\big|_{x^0=0}$$
$$= \partial_0 \int_{-\infty}^{\infty} \frac{\mathrm{d}p_3}{2\pi} \frac{\theta(|p_3| - m)}{\sqrt{|p_3|^2 - m|p_3|}} \sin\left(\sqrt{|p_3|^2 - m|p_3|}\, x^0\right) e^{ix^3 p_3} \bigg|_{x^0=0}$$
$$= \int_{-\infty}^{\infty} \frac{\mathrm{d}p_3}{2\pi} \theta(|p_3| - m)\, e^{ix^3 p_3} = \delta(x^3) - \frac{\sin mx^3}{\pi x^3}, \qquad (5.28)$$

which can be nonzero for finite $x^3$, as long as $m \neq 0$. (Recall that for purely spacelike $k_\mu = (0, 0, 0, 1)$ we have shown the ultra-locality of Eq. (5.13) above.) For the case of a purely timelike Chern–Simons parameter $k_\mu$, the commutator (5.26) thus violates microcausality.

# 6 Summary and discussion

Lorentz- and CPT-violating field theories might emerge as low-energy effective theories of a more fundamental theory, where Lorentz and CPT symmetry are broken spontaneously or dynamically [7]. Alternatively, these symmetry-breaking theories might result from a quantum anomaly within the realm of quantum field theory itself [4]. In both instances, the question arises whether or not these Lorentz- and CPT-violating theories are valid quantum field theories, that is, whether or not a consistent quantization is possible. For theories which contain an Abelian Chern–Simons-like term (1.2), issues like power-counting renormalizability and conservation of energy-momentum have already been discussed in Ref. [7], where it was demonstrated that these features continue to hold.

In this paper, we have focused on the issues of unitarity and causality for theories containing an Abelian Chern–Simons-like term (1.2). The results found strongly depend on whether the preferred direction $k_\mu$ of the Lagrangian term (1.2) is spacelike or timelike.

For a purely spacelike Chern–Simons parameter $k_\mu = (0, \vec{k})$, our results are certainly encouraging for the issue of quantization. By investigating the dispersion relations we have found in Section 2 that a universal, direction-independent signal propagation speed $c$ can still be defined (in this paper, we have chosen units such that $c = 1$). In addition, the group velocity is less or equal to $c$. This suggests that the CPT-violating term in Eq. (1.1) does not change the causal structure of spacetime, but rather acts like a medium with a direction-dependent dispersion for the field excitations (e.g. the photons).

In fact, the anisotropic propagation of the circular polarization modes makes clear that the Abelian Chern–Simons-like term for purely spacelike $k_\mu = (0, \vec{k})$ is T-odd (and P- and C-even). According to the dispersion relation (2.4) for Chern–Simons parameter $\vec{k} = (0, 0, 1)$, left-handed wave packets propagating in different directions along, say, the $x^2$-axis (with an infinitesimal $p_3$-component added) have unequal group velocities, $|\vec{v}_g^L(0, p_2, \delta p_3)| \neq |\vec{v}_g^L(0, -p_2, -\delta p_3)|$, as long as $m \neq 0$. The physics is thus non-invariant under time reversal, which flips the momentum, $\vec{p} \to -\vec{p}$, and preserves the helicity, $L \to L$, cf. Ref. [22].

We have also found from the dispersion relations that a separation into positive and negative frequency modes is still possible for purely spacelike $k_\mu$. Therefore, particles and antiparticles may be defined and the field may be quantized in the usual fashion [27]–[29]. As shown in Section 5.1, the resulting commutators of the electric and magnetic fields vanish for spacelike separations, which demonstrates that microcausality holds for the potential quantum field theory based on the Lagrangian (1.1). This result is not quite trivial. Recall, for example, that the CPT-violating theories of Ref. [13], with self-conjugate bosons of odd-half-integer isospin, fail precisely in this respect, because certain fields do not commute outside the light-cone. The well-known result [11] that microcausality, Lorentz invariance and the existence of a unique vacuum state imply CPT invariance does not contradict our result, because in the Lagrangian (1.1) Lorentz invariance is broken as well ($k_\mu$ is fixed once and for all).

In addition, we have demonstrated in Section 3 that a Feynman propagator for the relevant Lagrangian (1.1) can be defined in Minkowski spacetime with the usual $i\epsilon$ prescription and that this propagator can be Wick-rotated to Euclidean space. As shown in Section 4, reflection positivity holds for the Euclidean Feynman propagator. Again, this requires the choice of a (purely) spacelike $k_\mu$.

For both a classical and a quantum treatment, the causal structure of the Maxwell–Chern–Simons theory (1.1) thus remains unaltered by the inclusion of the CPT-violating term, provided that the Chern–Simons parameter $k_\mu$ in Eq. (1.2) is spacelike.

This suggests that a CPT- and Lorentz-symmetry-violating theory like (1.1) may lead to a consistent local quantum field theory. If so, the particular chiral gauge field theories discussed in Ref. [4], which display the remarkable phenomenon of a CPT anomaly, could perhaps be realized in nature (see also the remarks below).

On the other hand, a consistent quantization for a timelike Chern–Simons parameter $k_\mu$ does not seem to be possible. As noted in Section 3, the presence of imaginary energies at low momenta requires the exclusion of these momenta if unitarity is to be maintained.[7] But we have seen in Section 5.2 that this exclusion leads to a violation of microcausality. It is, therefore, not possible to maintain both unitarity and causality. In fact, these results are just the quantum analogs of the results of Ref. [6]. The authors of that paper have pointed out that the Green's functions of the classical equations of motion resulting from the Lagrangian (1.1) either are causal but with exponential growth in time, or without exponential growth but noncausal. For timelike $k_\mu$, there are no Green's functions that are both causal (i.e. propagating signals only into the future) and without exponential growth. As shown in Section 4 of this paper, reflection positivity in the Euclidean formulation is indeed violated for the Feynman propagator with Chern–Simons parameter $k_\mu = (1, 0, 0, 0)$.

Hence, the Abelian Chern–Simons-like term (1.2) for timelike $k_\mu$, which contains a P-odd (and C- and T-even) part, does not allow for a consistent quantization. It does appear, however, that a Chern–Simons-like term could play a role for T-violation (leaving C- and P-invariance intact), provided the parameter $k_\mu$ is (purely) spacelike. Such a Chern–Simons-like term would, in fact, provide a "fundamental arrow-of-time," cf. Ref. [1]. (Recall the Gedankenexperiment with the left-handed wave packet presented at the beginning of this section.) This problem is currently under investigation.

As briefly mentioned in Ref. [4], the birefringence of a photonic Chern–Simons-like term (1.2) with purely spacelike parameter $k_\mu$ could also affect the polarization of the cosmic microwave background. The expected polarization pattern [30] around temperature hot- and cold-spots would be modified, due to the action of the Chern–Simons-like term on the photons traveling between the last-scattering surface (redshift $z \sim 10^3$) and the detector ($z = 0$). Future satellite experiments such as NASA's Microwave Anisotropy Probe and ESA's Planck Surveyor could look for this effect, see Ref. [31] for further

---

[7]Moreover, tachyon pair production would destabilize the perturbative vacuum state, see Section 4 of Ref. [19].

details.

Let us end this section with three somewhat more theoretical issues. The first issue concerns stability, which has been discussed recently for a theory of a massive Dirac fermion with (spontaneous) Lorentz and CPT breaking [32]. For the Maxwell–Chern–Simons theory (1.1), the photon is, of course, stable. This holds, in particular, for the case of purely spacelike Chern–Simons parameter $k_\mu$. But even in the context of the CPT anomaly [4], with all chiral fermions integrated out, the effective action can be expected to have additional quartic (and higher-order) interaction terms for the photons. A rough estimate suggests an extremely small, but nonzero, effect for the decay of the photon, cf. Eq. (1.3). (The photons of the cosmic microwave background would not be affected significantly.) Whether or not photon decay is physically acceptable remains an open question, though.

The second issue concerns the case of a "lightlike" Chern–Simons parameter $k_\mu$, which has not been discussed so far. No imaginary energies appear for "lightlike" $k_\mu$. There is, therefore, no obvious obstacle against quantization, as was the case for timelike Chern–Simons parameters. But explicit calculations of the type performed in Sects. 4 and 5 are hampered by the complicated pole structure for the "lightlike" case. For the CPT anomaly, there may still be a problem with the "lightlike" theory, as will become clear shortly.

The third, and final, issue concerns the possible implications of our microcausality results for the case of an anomalous origin of the Lagrangian (1.1) considered here. As mentioned several times by now, it has been shown in Ref. [4] that certain chiral gauge field theories defined over the Euclidean spacetime manifold $M = \mathbb{R}^3 \times S^1$, for example, could give rise to a CPT anomaly of the form of a Chern–Simons-like term (1.2), with an additional factor $i$ for the Euclidean signature of the metric. (In the context of the CPT anomaly, the exponential of the integrated Euclidean Chern–Simons-like term appears as the phase factor of the chiral determinant, whereas the absolute value of the chiral determinant is CPT invariant.) In that case, the specific parameters of Eq. (1.2) would be $m = (2\,n + 1)\,\alpha/L$ and $k_\mu = (0, 0, 0, 1)$, with the integer $n$ defining the theory in the ultraviolet and the nonzero entry of $k_\mu$ corresponding to the single compact dimension of length $L$.

The microcausality results of the present paper then imply that this compact dimension should correspond to a spatial direction after the Wick rotation to the Lorentzian signature of the metric ($x^3 \in S^1$ becoming a spatial coordinate and $x^4 \in \mathbb{R}$, say, the time coordinate). If, on the other hand, the compact dimension were to correspond to the time direction ($x^4 \in S^1$), our results would lead us to expect problems with unitarity or causality. Indeed, we would then have started from a spacetime manifold with closed timelike curves, which has a built-in violation of what might be called "macrocausality," cf. Section 8.2 of Ref. [33].[8] Still, the proper fundamental theory (gauge fields and chiral fermions over a spacetime manifold with a separable compact dimension that is spacelike) and the corresponding effective theory (gauge fields with a Chern–Simons-like term for purely spacelike parameter $k_\mu$) appear to be consistent as far as causality is concerned.

## Acknowledgements

It is a pleasure to thank the anonymous referee for useful comments.

## A  Commutators for the Coulomb gauge

In this appendix, we first discuss the equations of motion of the Lagrangian (1.1) for purely spacelike Chern–Simons parameter $k_\mu = (0, \vec{k})$, but with the general axial gauge replaced by the Coulomb gauge $\vec{\partial} \cdot \vec{A} = 0$. From these equations of motion, we then determine the "tensor" structure $T_{ij}$ of the gauge field commutator (5.1). Finally, we calculate the commutators of the electric and magnetic fields.

The equations of motion in momentum space are (with the conventions $g_{\mu\nu} = \mathrm{diag}(-1,1,1,1)$, $\epsilon_{0123} = \epsilon_{123} = 1$, and using $\vec{p} \cdot \vec{A} = 0$)

$$(p_0^2 - |\vec{p}|^2)g^{\mu\nu} A_\nu - p^\mu p_0 A_0 + im\epsilon^{\mu\nu\rho l} A_\nu p_\rho k_l = 0 \ . \tag{A.1}$$

This leads to a nondynamical equation for $A_0$,

$$A_0 = i \frac{m}{|\vec{p}|^2} \epsilon_{abc} A_a p_b k_c \ , \tag{A.2}$$

---

[8] Another spacetime manifold with closed timelike curves would be, for example, $M = S^1 \times \mathbb{R}^2 \times S^1$, with periodic coordinates $x^4 \in [0, L_4]$ and $x^3 \in [0, L_3]$ and noncompact coordinates $x^1, x^2 \in \mathbb{R}$. The corresponding effective action could then contain a Chern–Simons-like term (1.2) with parameter $k_\mu \propto (L_4^{-1}, 0, 0, L_3^{-1})$. For $L_4 = L_3$ exactly, the Chern–Simons parameter $k_\mu$ would be "lightlike." Note also that the transformation of $k_\mu$ as given by Eq. (2.2) may not be applicable in general for this manifold (compare with Eq. (13) of Ref. [5]).

and dynamical equations for $A_i$,

$$M_{ij}A_j \equiv \left(\left[p_0^2 - |\vec{p}|^2 - m^2 \frac{|\vec{p}|^2|\vec{k}|^2 - (\vec{p}\cdot\vec{k})^2}{|\vec{p}|^2}\right]\delta_{ij}\right.$$
$$\left. - imp_0\left[\frac{p_i}{|\vec{p}|^2}\epsilon_{jab}p_a k_b + \epsilon_{ija}k_a\right] - m^2\frac{\vec{p}\cdot\vec{k}}{|\vec{p}|^2}p_i k_j + m^2 k_i k_j\right)A_j = 0\,. \quad \text{(A.3)}$$

The gauge field commutator (5.1) will obey the equations of motion (A.3) provided that

$$M_{ij}T_{jk}(p) = \pi_{ik}\,D^{-1}(p)\,, \quad \text{(A.4)}$$

with

$$D^{-1}(p) = (p^2)^2 + m^2\left(p^2|\vec{k}|^2 - (\vec{p}\cdot\vec{k})^2\right)\,. \quad \text{(A.5)}$$

(The tensor structure on the right-hand side of Eq. (A.4) could, in principle, be more complicated, but turns out to be just $\pi_{ik}$.) With some effort, it can be verified that (A.4) holds for the $T_{ij}$ as given by Eq. (5.3).

The magnetic fields then have the following commutator:

$$[B_i, B_j](p) = i^2 \epsilon_{iab}p_a \epsilon_{jcd}(-p_c)T_{bd}\,i\,D(p) = \Big((p_0^2 - |\vec{p}|^2)(\delta_{ij}|\vec{p}|^2 - p_i p_j)$$
$$+ imp_0(\vec{p}\cdot\vec{k})\epsilon_{ijl}p_l - m^2\{\delta_{ij}(|\vec{p}|^2|\vec{k}|^2 - (\vec{p}\cdot\vec{k})^2)$$
$$- p_i p_j|\vec{k}|^2 - k_i k_j|\vec{p}|^2 + (p_i k_j + k_i p_j)(\vec{p}\cdot\vec{k})\}\Big)\,i\,D(p)\,. \quad \text{(A.6)}$$

The electric fields are more involved,

$$\vec{E}(x) = \partial_0\vec{A}(x) - m\int d^3z\,\Delta^{-1}(\vec{x} - \vec{z})\,\vec{\partial}^z\epsilon_{abc}\,k_a\,\partial_b^z\,A_c(x^0, \vec{z})\,, \quad \text{(A.7)}$$

and one finds after some algebra the following commutators:

$$[E_i, B_j](p) = \Big((p_0^2 - |\vec{p}|^2)\epsilon_{ijl}p_l p_0 - m^2 p_0 k_i \epsilon_{jab}p_a k_b - imp_0^2(\vec{p}\cdot\vec{k})\delta_{ij}$$
$$+ im(\vec{p}\cdot\vec{k})p_i p_j + im(p_0^2 - |\vec{p}|^2)p_i k_j\Big)\,i\,D(p)\,, \quad \text{(A.8)}$$

$$[E_i, E_j](p) = \Big((p_0^2 - |\vec{p}|^2)\delta_{ij}p_0^2 - (p_0^2 - |\vec{p}|^2)p_i p_j$$
$$- m^2 p_0^2 k_i k_j + imp_0^3\epsilon_{ijl}k_l + imp_0(p_i\epsilon_{jab}p_a k_b - p_j\epsilon_{iab}p_a k_b)$$
$$- \frac{p_i p_j}{|\vec{p}|^2}\{(|\vec{p}|^2 - p_0^2)^2 + m^2((|\vec{p}|^2 - p_0^2)|\vec{k}|^2 - (\vec{p}\cdot\vec{k})^2)\}\Big)\,i\,D(p)\,. \quad \text{(A.9)}$$

Apparently, the electric field commutator contains a term with a $|\vec{p}|^{-2}$ pole. But this term is multiplied by precisely the function $D^{-1}(p)$, which cancels the over-all factor $D(p)$. Therefore, this term does not contribute to the contour integral (5.2), and we reproduce the commutators (5.7) – (5.9) as given in the main text.

# References

[1] R.M. Wald, Phys. Rev. D 21 (1980) 2742.

[2] S.W. Hawking, Phys. Rev. D 32 (1985) 2489.

[3] V.A. Kostelecký, R. Potting, Nucl. Phys. B 359 (1991) 545.

[4] F.R. Klinkhamer, Nucl. Phys. B 578 (2000) 277.

[5] F.R. Klinkhamer, J. Nishimura, Phys. Rev. D 63 (2001) 097701.

[6] S.M. Carroll, G.B. Field, R. Jackiw, Phys. Rev. D 41 (1990) 1231.

[7] D. Colladay, V.A. Kostelecký, Phys. Rev. D 55 (1997) 6760; D 58 (1998) 116002.

[8] S. Coleman, S.L. Glashow, Phys. Rev. D 59 (1999) 116008.

[9] R. Jackiw, V.A. Kostelecký, Phys. Rev. Lett. 82 (1999) 3572.

[10] M. Chaichian, W.F. Chen, R. Gonzales Felipe, Phys. Lett. B 503 (2001) 215.

[11] R. Jost, Helv. Phys. Acta 30 (1957) 409.

[12] R. Streater, A. Wightman, *PCT, Spin and Statistics, and All That*, Benjamin, New York, 1964.

[13] P. Carruthers, Phys. Rev. Lett. 18 (1967) 353; Phys. Lett. B 26 (1968) 158.

[14] S. Chern, J. Simons, Ann. Math. 99 (1974) 48.

[15] G. Leibbrandt, *Noncovariant gauges*, World Scientific, Singapore, 1994.

[16] B. Nodland, J.P. Ralston, Phys. Rev. Lett. 78 (1997) 3043.

[17] J.F.L. Wardle, R.A. Perley, M.H. Cohen, Phys. Rev. Lett. 79 (1997) 1801.

[18] L. Brillouin, *Wave Propagation and Group Velocity*, Academic, New York, 1960.

[19] A.A. Andrianov, R. Soldati, L. Sorbo, Phys. Rev. D 59 (1999) 025002.

[20] I. Montvay, G. Munster, *Quantum Fields on a Lattice*, Cambridge U.P., Cambridge, 1994.

[21] M. Veltman, *Diagrammatica–The Path to Feynman Rules*, Cambridge U.P., Cambridge, 1994.

[22] S. Weinberg, *The Quantum Theory of Fields I*, Cambridge U.P., Cambridge, 1996.

[23] J. Dhar, E. Sudarshan, Phys. Rev. 174 (1968) 1808.

[24] T. Jacobson, N.C. Tsamis, R.P. Woodard, Phys. Rev. D 38 (1988) 1823.

[25] K. Osterwalder, R. Schrader, Comm. Math. Phys. 31 (1973) 83; 42 (1975) 281.

[26] I.S. Gradshteyn, I.M. Ryzhik, *Table of Integrals, Series and Products*, Academic, New York, 1980.

[27] W. Heitler, *The Quantum Theory of Radiation*, 3rd ed., Oxford U.P., London, 1954.

[28] J.M. Jauch, F. Rohrlich, *The Theory of Photons and Electrons*, 2nd ed., Springer, New York, 1976.

[29] N.N. Bogoliubov, D.V. Shirkov, *Introduction to the Theory of Quantized Fields*, Wiley, New York, 1959.

[30] D. Coulson, R. Crittenden, N. Turok, Phys. Rev. Lett. 73 (1994) 2390.

[31] N.F. Lepora, arXiv:gr-qc/9812077; M. Kamionkowski, A. Kosowsky, Ann. Rev. Nucl. Part. Sci. 49 (1999) 77.

[32] V.A. Kostelecký, R. Lehnert, Phys. Rev. D 63 (2001) 065008.

[33] R.M. Wald, *General Relativity*, Chicago U.P., Chicago, 1984.

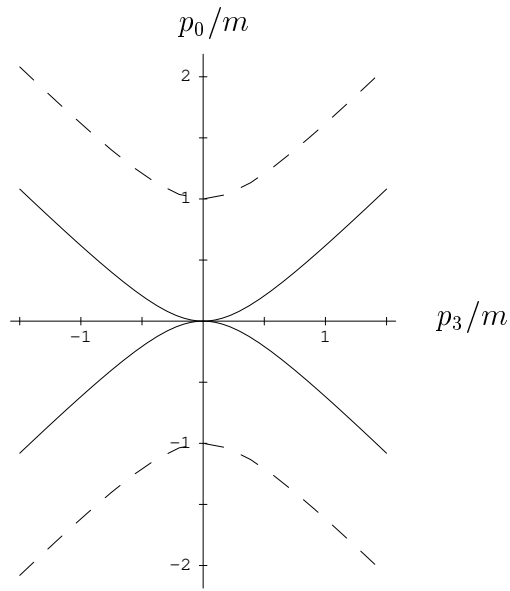

Figure 1: The dispersion relation (2.3) in the $(p_0, p_3)$ plane for a purely spacelike Chern–Simons parameter $k_\mu = (0,0,0,1)$ and mass scale $m$, with broken (solid) curves corresponding to the plus (minus) sign in Eq. (2.3).

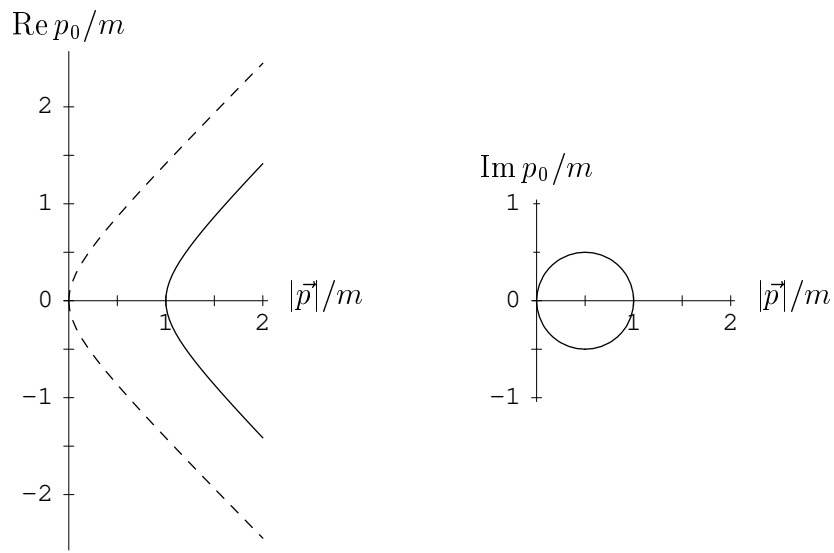

Figure 2: The dispersion relation (2.8) in the $(\operatorname{Re} p_0, |\vec{p}|)$ halfplane for a purely timelike Chern–Simons parameter $k_\mu = (1,0,0,0)$ and mass scale $m$, with broken (solid) curves corresponding to the plus (minus) sign in Eq. (2.8). For the minus sign in Eq. (2.8), the energy $p_0$ becomes imaginary at low momenta $|\vec{p}| < m$.